\documentclass[aps,epsfig,showpacs]{revtex4}
  
\usepackage{latexsym}  
\usepackage{graphicx}  
  
\begin{document}  
  
\title{On the asymmetry of Gamow-Teller $\beta $-decay rates in mirror nuclei  
 in relation with second-class currents}  
  
\author{N.~A.~Smirnova$^{1,2}$, C.~Volpe$^{3,4}$}  
  
\address{$^1$ Instituut voor Kern- en Stralingsfysica, University of Leuven,   
 Celestijnenlaan 200 D, 3001 Leuven, Belgium}  
  
\address{$^2$ University of Ghent, Institute for Nuclear Science,  
  Proeftuinstraat 86, B-9000 Ghent, Belgium; E-mail: nadya.smirnova@rug.ac.be}  
  
\address{$^3$ Institut f\"ur Theoretische Physik, Universit\"at Heidelberg,  
Philosophenweg 19, D-69120 Heidelberg, Germany;  
 E-mail: volpe@tphys.uni-heidelberg.de}  
  
\address{$^4$ Institut de Physique Nucl\'eaire, B\^at.~100, F 91406, Orsay cedex, France}  
  
\date{\today}  
\begin{abstract}  
The theoretical evaluation of major nuclear structure effects on  
  the asymmetry of allowed Gamow-Teller  
  $\beta $-decay rates in light mirror nuclei is presented.  
  The calculations are performed within the shell model,  
  using empirical isospin-nonconserving  
  interaction and realistic Woods-Saxon radial wave functions.  
  The revised treatment of $p$-shell nuclei is supplemented by systematic  
  calculations  
  for $sd$-shell nuclei and compared to experimental asymmetries when  
  available.  
  The results are important in connection with the possible existence  
  of second-class currents in the weak interaction.  
\end{abstract}  
  

\maketitle  
  
\vspace{1cm}

\noindent Corresponding author: \\ Nadya A. Smirnova\\ University of Ghent, Institute for Nuclear Science, \\  
 Proeftuinstraat 86, B-9000 Ghent, Belgium\\  
 Tel: (+32) (0)9 264 65 41  
\\ Fax: (+32) (0)9 264 66 97 \\ E-mail: nadya.smirnova@rug.ac.be \\

\section{Introduction}  
  
 One of the possibilities to look for the existence of second-class currents in  
 the weak interaction is through the study of differences in allowed Gamow-Teller (GT)  
 $\beta $-decay rates of mirror nuclei \cite{We58,HuGr63,KDR,Grenacs,Wi00}.

Weak processes in nuclei at low energy are well described by  
 the effective $\cal V-A$ interaction :  
 \begin{equation}\label{e1}  
 H_{\cal V-A}= \frac{G_F}{\sqrt{2}} J^{\dagger}_{\mu }j^{\mu }+ {\rm h.c.},  
 \end{equation}  
 where $J^{\dagger}_{\mu }$ ($j^{\mu }$) is the hadronic (leptonic) current,  
 and $G_F$ is the weak interaction coupling constant.  
 The most general Lorentz covariant form for the hadronic current reads  
 $J^{\dagger}_{\mu}=V_{\mu}+ A_{\mu}$ with  
\begin{eqnarray}  
V_{\mu} & = &i \bar{\psi}_p \left[g_V \gamma_{\mu} + {g_M \over  
    {2M}} \sigma_{\mu \nu} k_{\nu} + i g_S k_{\mu} \right] \psi_n \, ,  
     \label{e2} \\  
A_{\mu} & = &i \bar{\psi}_p \left[g_A \gamma_{\mu}\gamma_5 + {g_T \over  
    {2M}} \sigma_{\mu \nu}\gamma_5 k_{\nu} + i g_P  
    k_{\mu}\gamma_5 \label{e3}  
\right] \psi_n  
\end{eqnarray}  
 standing for the vector and axial-vector parts.  
 Here $k_{\mu}$ is the momentum transferred, $M$ is the nucleon mass,  
 $\psi_p$ ($\psi_n$) is the proton (neutron) field operator.  
 The six terms are the  
 main vector, weak magnetism, induced scalar in (\ref{e2}) and the main axial-vector,  
 induced tensor and induced pseudoscalar in (\ref{e3}),  
 with $g_V$, $g_M$, $g_S$, $g_A$, $g_T$, $g_P$ being the corresponding coupling constants.

The weak hadronic currents in Eqs.~(\ref{e2})--(\ref{e3})  
 can be classified~\cite{We58} according  
 to their transformation properties under the ${\cal G}$-parity operation,  
  ${\cal G}={\cal C} \exp(i \pi {\cal T}_2) $, i.e. the product of  
 charge conjugation ($C$) and rotation over $180^{\rm o}$ about the 2-axis  
 in isospin space.  
 Under this transformation, the main vector term, weak magnetism in $V_{\mu }$ (\ref{e2}) and  
 the main axial-vector and induced  
  pseudoscalar terms in $A_{\mu }$ (\ref{e3}) are  
 even (${\cal G}V^I_{\mu } {\cal G}^{\dagger }=V^I_{\mu }$) and  
 odd ($ {\cal G}A^I_{\mu } {\cal G}^{\dagger }=-A^I_{\mu }$), respectively.  
 They are usually called first-class currents.  
 The induced scalar term in $V_{\mu}$ and induced tensor  
  in $A_{\mu}$ transform, under ${\cal G}$,  
  with opposite sign with respect to  
 first-class currents.  
 They are usually referred to as second-class currents ({\it scc}).

The interest in the related weak form factors stem both from the desire of understanding  
  them in the framework on quantum chromodynamics and in the search for  
   the physics beyond the Standard Model \cite{Mukho}.

Constraints on weak form factors come from  
two hypothesis that follow from approximate symmetries of the Standard Model :  
the conserved vector current (CVC) hypothesis  
and the partially conserved axial-vector current (PCAC) hypothesis.  
In the limit of exact isospin symmetry,  
the CVC hypothesis ensures that : i) the weak form factors  
$g_V(k^2)$ and $g_M(k^2) $ are related  
to the electromagnetic ones;  
ii) the induced scalar form factor $g_S(k)$ vanishes.
Therefore, the latter term can contribute only at the level of isospin  
symmetry breaking.  
  
 The PCAC hypothesis does not put any limits on the value of {\it scc}  
 induced tensor form factor $g_T(k^2)$. Thus, the non-conservation of the  
 ${\cal G}$ symmetry (due for example to differences in $u$ and $d$ quark  
 masses and to the electromagnetic interaction) will result in a non-zero  
 contribution of $g_T$.

 It is an experimental and theoretical challenge to place limits on the possible presence of  
 the induced tensor term in the axial-vector current.  
 The search for evidence of its existence is longstanding  
 \cite{Grenacs,Wi00}.  
  
In particle physics, the evidence can be obtained, for example, from $G$-parity non-conserving  
 decay branch, such as $\tau \rightarrow \eta \pi \nu$ \cite{Ba96}, or from the angular distribution  
 analysis after the decay $\tau \rightarrow \omega \pi \nu$ involving both first and second-class currents  
 \cite{omega}.  

 The study of weak processes in nuclei offers  
 various possibilities in this search.  
 There are two limits on the value of $g_T$  
 obtained from muon capture experiments, which are consistent with  
 the absence of second-class currents \cite{Hol,Gov}.

 In $\beta $ decay, one can determine possible contributions from the  
 induced tensor term in two types of measurements \cite{Grenacs}.  
 The first one is through correlation experiments  
 where one measures the correlation between nuclear spin and  
 momentum of emitted $\beta $-particle in the decay of oriented systems, or  
 the correlation between the momentum of the emitted $\beta $-particle and of the  
 radiation from a daughter state.  
 A recent re-measurement of $\beta $-particles angular distribution in the decay of  
 aligned $^{12}$B and $^{12}$N gives a non-zero value of $g_T$~\cite{Minami}.  
  These experiments require the knowledge  
 of the weak magnetism coupling constant, which gives another uncertainty.  
 Only in the last years an independent way to extract  
 {\it scc}, based on the $\beta $-$\alpha $ angular  
 correlation in $A=8$ system, has been proposed~\cite{DeBr,AmBe97}.

Besides, the second-class currents can give rise to different rates of mirror  
 $\beta _{\pm }$-decay transitions~\cite{We58}.  
 This phenomenon was actively explored in early seventies, stimulated by  
 first experimental observations of large asymmetries in the $ft$-values  
 in the decay of light nuclei~\cite{Wi70}.  
 The experimental asymmetry deduced from measured halflives  
 \begin{equation}  
 \label{delta1}  
 \delta =\frac{(ft)_+}{(ft)_-}-1 \; ,  
 \end{equation}  
 where $(ft)_{\pm }$ refer to the $\beta_{\pm }$ decays in the mirror nuclei,  
 is known at present to vary on average from $1 \%$ to several tens of  
 percent,  being about $(5\pm 4)$\% on average~\cite{comp}.  
 From the theoretical point of view, the asymmetry $\delta $ (\ref{delta1})  
 can have two origins :  
 i) the possible existence of the $scc$ term in the axial-vector current, i.e. non-zero  
 contribution of $g_T$ ($\delta^{scc}$);  
 ii) the breaking of the only approximate  
 isospin symmetry between the decaying nuclear states  
 ($\delta^{nucl}$), i.e.  
 \begin{equation}  
 \label{delta2}  
 \delta =\delta^{scc}+\delta^{nucl} \; .  
 \end{equation}

 Assuming the perfect mirror symmetry for the nuclear states and impulse approximation,  
 the induced tensor current gives rise~\cite{HuGr63} to the asymmetry $\delta^{scc}$  proportional  
 to the induced tensor constant  
 and the sum of maximum energy releases in both mirror transitions,  
 ($W^+_0+W^-_0$).  
 The more refined calculations~\cite{KDR},  
 including the off mass-shell effects and meson exchange corrections  
 which are crucial in many-body physics~\cite{Li71}, predict the asymmetry provided by  
 the second-class currents to be  
 \begin{equation}  
 \label{dscc}  
 \delta^{scc}=-4 \frac{\lambda }{g_A} J +  
 \frac{2}{3 g_A} (\lambda  L - 2 \zeta )(W^+_0+W^-_0).  
 \end{equation}  
 Here $\lambda $ and $\zeta $ are two parameters related to  
 the coupling constants for off-shell and mesonic corrections,  
 and induced tensor coupling constants,  
 $J$ and $L$ are ratios of the matrix elements of  
 meson-exchange currents between nuclear states (see Ref.~\cite{KDR} for more details).

However,  it has been shown~\cite{BSRo65,Bl71,Wi71,To73},  
 that the nuclear  structure effects provide the principal contribution to the  
 asymmetry, from 10\% to 20\% for different nuclei, making the extraction of  
 the possible contribution from induced tensor term an extremely difficult task~\cite{KDR,Wi00}.

All early estimations~\cite{BSRo65,Bl71,Wi71,To73} of $\delta^{nucl}$  
 were carried out within the nuclear shell model for a few $p$-shell nuclei only  
 (namely, for $A=8,9,12,13$),  
 since the proper treatment of the $sd$-shell nuclei was far outside available  
 at that time computational power.  
 The basic ingredients of the calculations were  
 the Cohen-Kurath interaction~\cite{CoKu65}  
 with empirical isospin-nonconserving (INC) corrections~\cite{To73}  
 and the radial wave functions obtained with the Woods-Saxon (WS) potential.  
 The results obtained have been used to extract limits on second-class  
 currents~\cite{KDR,Wi00}.

To be precise, the main contribution to $\delta^{nucl}$ comes from the difference in  
 the matrix elements of the GT operator ($\sigma \tau $) between nuclear  
 states, due to isospin mixing and differences in the radial wave  
 functions,  as we extensively discuss below.  
 The calculations of resulting asymmetry in $ft$-values  
 performed by different authors~\cite{BSRo65,Bl71,Wi71,To73}  
 vary significantly, either because of different choices in the parametrizations  
 or because of the more and more  
 refined shell model technique used by the subsequent authors.

 Apart from this, there are some contributions to $\delta^{nucl}$  
 due to the higher order effects,  
 most of which are usually incorporated as corrections to the statistical rate function $f$,  
 i.e. forbidden matrix elements,  
 known induced currents, such as the induced pseudoscalar  
 (see Ref.~\cite{HuGr63} for a complete list).  
 Contrary to the calculation of the main part of $\delta $,  
 the evaluation of higher order corrections,  
 using  different nuclear wave functions~\cite{BSRo65,To73},  
 gives quite similar values and in any case does not exceed a few percent.

Since the mentioned above estimations of the GT transition rates in mirror nuclei,  
 no other attempts to refine the calculations have been undertaken up to now,  
 in spite of the essential improvement in the knowledge of the low-energy  
 nuclear structure.  
 Wider and more accurate experimental data  
 on $\beta $ decay ~\cite{comp} are available nowadays,  
 including  $sd$-shell nuclei.  In Fig.~1 we show the asymmetry extracted from the measured  
 $\beta $-decay lifetimes of the selected mirror transitions for most of which the asymmetry is  
 different from zero by $2\sigma $ or more.  

 Besides, the theoretical description of $p$-shell and $sd$-shell nuclei has  
 made much progress during  
 the last thirty years~\cite{BrWi88}.  
 Since the limits on the induced tensor current are still not  
 settled~\cite{Wi00,Hol,Gov,Mukho,Minami},  
 it is worthwhile  to improve the constraints on second-class currents  
 coming from the $\beta$ decay of nuclei using on one hand the current experimental  
 systematics and on the other hand improved calculations.

The aim of this paper is to present up-to-date shell model calculations,  
 for $p$-shell and, for the first time, $sd$-shell nuclei, of the main nuclear structure effects on the asymmetry  
 $\delta^{nucl}$ (\ref{delta2}) arising from the difference of GT operator matrix elements due to  
 isospin mixing in nuclear states and differences in the radial wave functions.  
 The higher order corrections can present a subject for a future work.  
 In this study, we use $p$-shell~\cite{CoKu65}  
 and $sd$-shell model wave functions~\cite{usd} obtained from  
 the INC Hamiltonians~\cite{OrBr89}, in a sufficiently large model space easily  
 accessible now, and realistic radial dependence, and we compare the results with the experimental data.

 The theoretical framework, we choose,  follows the one of Towner~\cite{To73} and  
 goes beyond it in several respects. First, we use  
 the better fitted INC interaction to take into account isospin  
 breaking, both without and within the parent state formalism,  
 similarly to the calculations of  
 Ormand and Brown~\cite{OrBr85} for superallowed transitions.  
 Second, we use a different parametrization of the WS potential.  
 We carefully investigate the sensitivity of the asymmetry  
 to the way the tail of the single-particle  
 wave functions may be determined by fitting the  
 single-particle energies to the experimental  
 separation energies.  
In fact, there are two possible procedures : I) to adjust  the energy of the last  
  unoccupied single-particle state  only, as in Ref.~\cite{Wi71,To73}; II)  
 to fix all the energies of the single-particle states in the considered shell  
 at the experimental separation energy, as it is done in Ref.~\cite{Bl71} for $A=12$ only. Strictly  
 speaking, none of them is exact.  
 However, we will argue that the latter is more consistent with the  
 shell model, while the former can even lead to ``unrealistic'' results  
 when the last unoccupied single-particle state is unbound.  
 We will then use, contrary to previous calculations  
 \cite{Wi71,To73}, the procedure of fitting all the single-particle energies at  
 the same experimental separation energy to estimate the asymmetry for an ensemble of $p$- and  
 $sd$-shell nuclei. We fit the separation energy either  
 by adding a surface term to the WS potential  
 or by varying the volume term, as it is done in  
 Ref.~\cite{Bl71,Wi71,To73}.

In previous works on asymmetries and second-class currents~\cite{BSRo65,Bl71,Wi71,To73},  
it was never stressed that one should discuss the $B$(GT)-values before considering 
the corresponding asymmetries. Here we carefully examine 
the $B$(GT)-values and test the wave functions by comparing certain electromagnetic observables,  
 namely magnetic and quadrupole moments, to available experimental data.  
 The problem of the effective GT transition operator in relation with the  
 quenching of the GT strength is also discussed.

 We remark that to extract limits on second-class currents in  
 $\beta$ decay one needs calculations on $B$(GT)-values as precise as possible and  
 therefore should concentrate on the best described nuclei.  
 In the context of the shell model, it means that  
 very light nuclei, possibly viewed as few-nucleon systems, may in principle not be  
 very good candidates.  
For example, the lightest nuclei, such as those involved in the $A=8,9$ cases,  
should be treated with particular techniques.  
 This is also why, contrary to what was done in the past,  
in the present work we make an extensive analysis of both $p$-shell and $sd$-shell nuclei (Fig.~1).

In the present study we restrict ourselves to the WS wave  functions.  
 As was noticed in Ref.~\cite{OrBr85}, the WS potential may overestimate the difference in  
 the tails of proton and neutron radial wave functions due to  
 differences in the Coulomb and isovector potentials.  
The results shown in \cite{OrBr85} indicate that  
 calculations with the wave functions obtained from the  
 Hartree-Fock procedure would lead to a smaller effect on the asymmetry.  
However, as we will show, the present level of accuracy on the $B$(GT)-values  
is still not good enough for extracting possible contributions  
 of the second-class currents from the asymmetries.  
In this respect, calculations with Hartree-Fock wave functions would  
not improve the $B$(GT) estimates enough to change these  
 conclusions. For these reasons, we present results obtained with WS  
 wave functions only which provide the upper limit of theoretical asymmetries.

 The paper is organized as follows. In Section II we describe the  
 theoretical approach, we give the details of the interaction used,  
 the parameters of the single-particle WS potential and discuss the different procedures.  
 Section III is devoted to the presentation of the results  
 on magnetic and quadrupole moments, transition probabilities and asymmetries  
 obtained for $p$- and $sd$-shell nuclei.  
 Summary and conclusions are given in the last section.

\section{Theoretical framework}

 Within the nuclear shell model~\cite{BrGl77},  
 the asymmetry (\ref{delta1}) is related to the GT matrix elements  
 $M_{\pm }$ for  $\beta_+$ and $\beta_-$ decay through  
\begin{equation}  
\label{delta}  
 \delta = \left|\frac{M_-}{M_+}\right|^2 -1 \ ,  
\end{equation}  
 with  
 \renewcommand{\arraystretch}{1.5}  
 \begin{equation}  
 \label{MGT}  
 \begin{array}{lcl}  
 \displaystyle M_{\pm } & \equiv &  
 \langle f || \sum_{k=1}^{n} \hat \sigma(k) \hat \tau_{\pm } (k) || i \rangle = \\  
 \displaystyle  
  & = & \sum\limits_{j_1,j_2, \pi } (-1)^{J_f+J_{\pi }+j_2+1} \sqrt{(2J_i+1)(2 J_f+1)}  
\left\{  
 \begin{array}{ccc}  
 J_f & J_i & 1 \\ j_1 & j_2 & J_{\pi } \\  
 \end{array}  
 \right\} \\  
 \displaystyle  
  & & S^{1/2}(j_2;J_f J_{\pi }) \, S^{1/2}(j_1;J_i J_{\pi })  
  \langle n_2 l_2 j_2 || \hat \sigma \hat \tau_{\pm } || n_1 l_1 j_1 \rangle \; ,  
 \end{array}  
 \end{equation}  
 where $n$ is the number of valence particles in the many-body wave  
 functions that  
 describe the initial, $|i \rangle $, and final, $|f \rangle $,  
 states in mother and daughter nuclei with angular momenta  
 $J_i$ and $J_f$ respectively;  
 $J_{\pi }$ is the angular momentum of the parent states $\pi$ of the  
 $(A-1)$ nucleus and  
 $S^{1/2}(j;J J_{\pi })$ denotes a spectroscopic amplitude with a proper phase factor, and 
 gives the mixing of different shell model configurations in the eigenfunction.  
 The INC terms in the effective shell-model interaction are responsible for the difference in the values of  
 $S^{1/2}(j;J J_{\pi })$ for mirror transitions.  
 The single-nucleon reduced matrix element is defined as  
 \begin{equation}  
 \label{spme}  
  \langle n_2 l_2 j_2 || \hat \sigma \hat \tau_{\pm } || n_1 l_1 j_1 \rangle =  
  \delta_{l_1 l_2} (-1)^{j_1+l_1+3/2} \sqrt{6} \sqrt{(2j_1+1)(2j_2+1)}  
 \left\{  
 \begin{array}{ccc}  
 1/2 & j_2 & l_1 \\ j_1 & 1/2 & 1 \\  
 \end{array}  
 \right\}  
 \Omega ^{\pi }_{j_1 j_2} \; ,  
 \end{equation}  
 with $(n_i,l_i,j_i)$ labelling the single-particle states in a spherical basis and  
 $\Omega ^{\pi }_{j_1 j_2}$ being the overlaps of the single-nucleon  
 radial wave functions :  
 \begin{equation}  
 \label{rad}  
 \Omega ^{\pi }_{j_1 j_2}=\int_{0}^{\infty }  
 R^{\pi }_{n_1 l_1 j_1}(r)  R^{\pi }_{n_2 l_2 j_2}(r) r^2 dr  \; .  
 \end{equation}  
 The radial single-particle wave functions $R^{\pi }_{n_1 l_1 j_1}(r)$ are obtained from  
 a spherically symmetric single-particle potential.  
 For realistic radial wave functions, obtained from WS or Hartree-Fock  
 potential with Coulomb and other charge-dependent  
 corrections, the integrals (\ref{rad}) are non-identical for two mirror processes.  
 The difference in the matrix elements of GT operator for $\beta_+$ and  
 $\beta_-$-decays (\ref{MGT}) is extremely sensitive to the asymptotics of  
 the radial part of the single-nucleon wave functions  
 which depends on the particle separation energy through  
 \begin{equation}  
\label{asymp}  
 R(r) \propto \exp{\left(-\sqrt{\frac{2m |E - E_{\pi }|}{\hbar^2}}r\right)} \; ,  
 \end{equation}  
 where $E$ and $E_{\pi }$ are the energies of the initial  
 state of the $A$-nucleus and of the final state in the  
  parent $(A-1)$-nucleus, respectively, and $m$ is the reduced nucleon mass~\cite{Bl71}.  
 \par  
 In case the number of the parent states to be taken into account  in $(A-1)$-nucleus is too large,  
 one is obliged to introduce different approximations. 
There exist two basic possibilities to deal with the problem:  
 either to truncate the number of states, leaving only those  
 which mostly contribute to the matrix element (\ref{MGT}), and then scale the result, or  
 to neglect completely the dependence on $\pi $ and approximate all integrals by those corresponding  
 to the $(A-1)$ nucleus being in its ground state.  
 We have used the latter procedure.  
 In this case, the expression (\ref{MGT}) reduces to  
\begin{equation}  
 \label{MGT5}  
 M_{\pm } = \sum_{j_1,j_2} {\rm OBTD}(j_1,j_2; \Delta J=1)  
  \langle n_2 l_2 j_2 || \hat \sigma \hat \tau_{\pm } || n_1 l_1 j_1\rangle \; ,  
 \end{equation}  
 where OBTD$(j_1,j_2; \Delta J=1)$ are one-body transition densities~\cite{OrBr85}.

We will present the results on $B$(GT)-values obtained using both Eq.~(\ref{MGT5}) and  
 Eq.~(\ref{MGT}), where we have taken into account up to as many as 100 states of a parent nucleus.  
 Provided this number is still too small for a good approximation (as it is for some $sd$-shell nuclei  
 for which the dimensions of the eigenvalue problem reaches a few thousands),  
 we have restricted ourselves only to the calculations  
 according to Eq.~(\ref{MGT5}).

 To summarize, within the shell model the asymmetry  
 in the decay of mirror nuclei can have two origins:  
 i) inequivalent $S^{1/2}$ values in (\ref{MGT}) or OBTD's in (\ref{MGT5})  
 for $\beta_+$ and $\beta_-$ decays due to isospin breaking effects;  
 ii) difference of the overlaps between the proton and neutron  
 radial wave functions (\ref{rad}) for mirror transitions.  
 (These two nuclear structure effects were referred to, in the  
 past, as effects due to Coulomb and charge-dependent nuclear forces and  
 binding energy phenomena, respectively \cite{To73}.)  
The two factors can be estimated independently and  
therefore we will discuss them  separately.

\subsection{Isospin corrections to configuration mixing }  
  
 The isospin symmetry in nuclei is only an approximate symmetry due to  
 the presence of the Coulomb interaction and isospin non-conserving nuclear forces.  
 Within the shell model, both effects can be well incorporated in a phenomenological way.  
 In Ref.~\cite{OrBr89}, the INC Hamiltonians have been derived by supplementing several  
 empirical shell model interactions (e.g., Cohen-Kurath for $p$-shell nuclei,  
 the interaction of Zuker and co-workers~\cite{ZBM} within the  
 $(p_{1/2}d_{5/2}s_{1/2})$ shell model space  for nuclei  
 around $^{16}$O, the interaction of Wildenthal~\cite{usd} for $sd$-shell nuclei) by extra terms  
 including isospin non-conserving ones.  
These are isovector single-particle energies for a given shell, matrix elements of  
 Coulomb force and of the isovector and isotensor parts of nucleon INC interaction,  
 whose strengths were fitted to the parameters of the isobaric mass-multiplet equation.  
  The interaction yields the shifts of the levels in mirror  
 partners often within 100 keV precision compared to experimental ones.  
 We adopt these INC interactions for all our calculations within  
 the $p$, $(p_{1/2}d_{5/2}s_{1/2})$ and  $sd$ configuration spaces.

\subsection{Radial wave functions}  
  
 We determine the single-particle radial wave function $R_{nlj}(r)$ of a nucleon  
 participating in the $\beta$ decay  
 by solving the Schr\"odinger equation with a WS potential :  
\begin{equation}  
V(r)= -V_{ws}f(r) -V_{so}{{r^2_0} \over {r}}{{d} \over  
  {dr}}[f(r)]{\bf l} \cdot {\bf s} + V_{c} h(r)  
\label{SW}  
\end{equation}  
 where  
\begin{eqnarray}  
 \displaystyle f(r) &  = & \frac{1}{\left[ 1+ \exp{(\frac{r-R_0}{a})}\right]} \\  
 \displaystyle   \nonumber h(r) & = &  
 \left\{  
 \begin{array}{l}  
 \displaystyle   \frac1r \quad  {\rm for} \quad r \ge R_0 \\  
 \displaystyle   \frac{1}{2R_0}\left(3 - {r^2 \over R_0^2}\right) \quad  {\rm for} \quad r < R_0 \\  
 \end{array}  
 \right.  
 \nonumber  
 \label{SWpar}  
 \end{eqnarray}  
 with $R_0=r_0(A-1)^{1/3}$ and $V_c=Ze^2$. The parameters are chosen  
 following the Bohr-Mottelson parametrization \cite{BM}, namely  
 $r_0=1.27$ fm, $a=0.75$ fm, $V_{ws}= V_{0}-V_{NZ}*(N-Z)t_z/A$, $V_{so}=V_{ls}*V_{ws}$,  
 $V_0=50.5$ MeV, $V_{NZ}=32$ MeV, $V_{ls}=0.22$.  
 $N$, $Z$, $A$ are the neutron, proton and mass numbers of the nucleus.  
  We have checked that this parametrization reproduces well charge radii.  
 The agreement for $p$-shell nuclei is within 5\%, except  
 for the halo nucleus $^{8}$B.  
 The WS potential used in Ref.~\cite{Bl71} does not contain any symmetry term, while  
 the one of Ref.~\cite{To73} differs in the parametrization  
 and includes an additional Coulomb term.  
  
 Remark that the potentials we use for the mirror partners differ only for the  
 Coulomb term.

 In general, there are two possible procedures  
 to  determine the radial part of the wave functions  
 by adjusting the energies of the single-particle states in the potential of the parent  
 $(A-1)$-nucleus to the experimental proton or neutron separation energy in the $A$-nucleus.  
 Method I : we fit the last unoccupied single-particle state and  
 keep the same potential to get all the other radial wave functions.  
 This choice was followed in Refs.~\cite{Wi71,To73}.  
 Method II : we adjust each single-particle energy at the separation energy of the nucleus,  
 as was done in Ref.~\cite{Bl71} for the $A=12$ case only.  
 We believe that the latter procedure is more consistent with the shell model  
 than the former because method I corresponds to the extreme independent particle model.  
 This conclusion is also supported by the unrealistic values for  
 the electromagnetic moments in $A=8,9$ obtained with method I.

In practice, the adjustment of the energies is performed either by including a surface term  
 ($\sim [f'(r)]^2$) or by varying the depth of the volume term in the potential.  
 We use the same set of parameters fitted for a given nucleus for its mirror partner. This automatically  
 yields single-particle energies very close to the experimental separation energies  
 (the difference does not exceed a few hundred keV and in some cases $1~$MeV).

 We have studied the sensitivity of the asymmetries to both methods I and II  
 (either including the surface term or by varying the depth of the volume term)  
 for $p$-shell nuclei as trial case, and  
 we have applied  method II only, the most consistent one,  
 to calculate the asymmetries for the $sd$-shell nuclei.

\section{Results}

 In this section we present the calculated magnetic and quadrupole  
 moments, the transition rates and the asymmetries for an ensemble of  
 $p$-shell and $sd$-shell nuclei.  
 The $S^{1/2}$ values and OBTD's have been obtained from  
 the shell model code Oxbash~\cite{Oxbash}.  
 We use different types of radial wave functions : Harmonic  
 Oscillator  (HO) and  
 Woods-Saxon (WS) wave functions obtained from the different fitting procedures as explained  
 in the previous section (method I and II), and  
 either neglecting the dependence on the parent state (\ref{MGT5})  
 or including it (\ref{MGT}--\ref{rad}) .  
 We also discuss how the results vary when one adjusts the single-particle  
 energies either by including a surface (S) or by changing the volume (V)  
 term of the WS potential.  
  
  The experimental $B$(GT)-values are obtained following the procedure outlined in  
 Ref.~\cite{BrWi85,ChWa93} with updated data on $Q$-values, half-lives and  
 branching ratios as specified in Tables II and III  
 (from the  compilation of asymmetries of allowed transitions ($\log(ft)<6$, $\pi_i=\pi_f$) in $A<40$ nuclei   
 presented in Ref.~\cite{comp}).  
 For the decay to unbound states in the A=8,9,17 systems, we use the approximation of  
 widthless levels and therefore we do not adopt the f-factors of Ref.~\cite{ChWa93}.   
  
 For comparison with theory, we have selected only the transitions for which  
 the error on the asymmetry is small enough, since our final purpose is to  
 extract the contribution of the second-class currents, if any.   
  
 Before discussing the results on $B$(GT)-values,  we comment on the problem of quenching of the  
 GT strength.  We also show what level of accuracy of the nuclear wave functions  
 we have, by considering the electromagnetic properties of some of the  
 nuclear states involved in the studied $\beta $-decay branches.

 \subsection{Effective GT-operator}

 It is was suggested long ago that the axial-vector coupling constant $g_A$ in GT
 transitions is quenched in the nuclear medium compared to the free space
 value (for review see, e.g., Ref.~\cite{Ost92} and references therein). 
 In fact, the experimentally measured total GT strength from charge-exchange
 reactions is largely suppressed compared to the model-independent sum rule~\cite{Ost92}, 
 as well as the $\beta $-decay transition rates are systematically overestimated
 by the $0\hbar \omega $ shell-model calculations (e.g., Ref.~\cite{BrWi85}). 
 The reasons for this discrepancy are 
 (i) lack of high-order configuration mixing, and 
 (ii) inadequacy of the one-body GT operator and need of taking into account meson-exchange
 currents, in particular, $\Delta $-isobar admixtures (the former effect is
 estimated to provide a dominant contribution to the quenching 
 within the $0\hbar \omega $ shell model~\cite{BrWi87,ToAr}).

%
%

 However, it is also well established~\cite{Wi7374,BrWi85}  
 that in the shell model, up to a good approximation, 
 the phenomenon can be taken into account by an empirical multiplicative factor,  
 $ \left(g_A/g_V\right)_{\rm eff}= \lambda \left(g_A/g_V\right)_{\rm bare} $  
 with a bare value for the ratio of 1.26 as obtained from neutron decay
 (see Ref.~\cite{Brown92} for certain cautions).  
 In this case, the quenching does not influence  
 asymmetries of the transition rates, i.e. the ratios of the matrix elements, which is then simply  
 given by (\ref{delta}).  
  
 Note, that here we are concerned with individual transitions, which   
 represent only a small fraction of the total GT strength, even   
 being the main branches of the $\beta $ decay.  
 Their analysis is more challenging because of the extreme sensitivity to configuration  
 mixing predictions, in particular, for weak transitions or in the  
 presence of close-lying final states.   
 Thus, the quenching phenomenon as described above can be more clearly seen  
 for stronger transitions.  
  
  Bearing these remarks in mind, in the sections below   
 we give the theoretical values for the GT-transition rates obtained with  
 $ \left(g_A/g_V\right)_{\rm bare}=1.26$.

 \subsection{Test of the wave functions}  
  
 The accuracy of the wave functions is one of the most critical issues of the study,  
 especially since we are interested in the fine details of the weak interaction.  
 To check the wave functions of the states involved in the $\beta $ decay considered below  
 and to demonstrate the present level of accuracy, we have computed  
 the transition probabilities of electromagnetic processes for some of the states of interest.  
 The electromagnetic processes are described up to an uncertainty arising mainly  
 from two factors, provided the best wave functions of the many-body problem are available:  
 i) the restricted model space used in the calculations; and ii) impulse  
 approximations for the operators.

 We have calculated some electromagnetic characteristics,  
 namely, magnetic dipole and electric quadrupole moments for the ground states,  
 which provide the most stringent test for the wave function of a given state.  
 We will follow the conventional way and use  
 the standard form of the M1 operator with the free-nucleon $g$-factors  
 ($g_s(\pi )=5.586$, $g_l(\pi )=1.0$, $g_s(\nu )=-3.826$, $g_l(\nu )=0.0$),  
 which gives in general a reasonable agreement with the data.  
 The electric quadrupole moments have been calculated with three different choices of effective  
 charges~\cite{BrWi82,BrWi88}.  
 The results are summarized in Table I and Fig.~2(a,b), where  
 the calculated values for certain states in $p$-shell nuclei  
 exploited below in the $\beta $-decay analysis are compared with the data.


 Both types of moments were computed according to  
 \begin{equation}  
 \label{em}  
 M(L) = \sum_{j_1,j_2} {\rm OBTD}(j_1,j_2; \Delta J=L)  
 \langle n_2 l_2 j_2 || \hat O_{L} || n_1 l_1 j_1 \rangle \; ,  
 \end{equation}  
 where $\hat O_{L}$ is magnetic dipole or electric quadrupole operator.  
 The OBTD values have been obtained from INC interaction, while the radial  
 wave functions from either the harmonic oscillator or the WS potential,  
 adjusted to give the separation energies only with respect to the ground state.

 The magnetic dipole moments are sensitive to the single-particle configurations and  
 almost do not depend on the radial form of the basis wave functions.  
 As seen from Table I and Fig.~2, the theoretical description of the magnetic moments  
 is always within 20\%, thus showing the high quality of the effective interaction we use here.

 The quadrupole moments depend on both the configurations involved and the radial  
 wave functions. The results obtained with both harmonic oscillator and WS radial wave functions  
 (method II, V) for three choices of the effective electric charges are given in Table I.  
 The calculations obtained  by fitting the surface term (S) gives very similar  
 results and to avoid repetition we do not present them.  
  
 To get the radial wave functions,  
 we have also tested the other possibility of adjusting the potential, namely,  
 fitting the separation energies  to the last unoccupied  nucleon state  
 (method I, V or S). We have found out that this procedure yields unrealistic  
 values of the quadrupole moments for light nuclei.  
 For example, the moments of $^8$Li calculated with the radial wave functions I V and I S  
 are $-0.85$ e.fm$^2$ and $-0.88$ e.fm$^2$ respectively, to be compared with the experimental  
 values and theoretical results obtained with method II (Table I).  
 We will see that the calculation of $\beta $-decay transition rates  
 does not help in choosing between method I and II,  
 because they both yield similar agreement with the measured $B$(GT)-values.  
 We emphasize that only method I (V or S) was used in the  
 past~\cite{Wi71,To73}, the case $A=12$ being the only exception \cite{Bl71}.

As seen from Table I,  
 both harmonic oscillator and WS radial wave functions (method II)  
 yield similar values of the quadrupole moments and on average reproduce  
 fairly well the experimental data.  
 It is difficult to use the results as an additional constraint for the potential,  
 since the quadrupole moments  are also rather sensitive to the choice of the effective charges  
 (see Table I).  
 The values obtained with the WS wave functions and the last set of the effective charges  
 ($e(\pi )=1.35$, $e(\nu )=0.35$)  
 are plotted in Fig.~2(b) and as seen agree within 20\% with the results of  
 the measurements.

 \subsection{GT-transition rates}

The calculated $B$(GT)-values,  
\begin{equation}  
 \label{BGT}  
 B({\rm GT}_{\pm})=\frac{(g_A/g_V)^2}{2J_i+1} \left|M_{\pm }\right|^2 \; ,  
 \end{equation}  
 are summarized in Tables II, for $p$-shell nuclei, and III, for $sd$-shell nuclei,  
 in comparison with the experimental data (Fig.~3).  
 The theoretical results are organized into columns labelled by the type of calculations.  
 Contrary to the moments discussed above,  
 the $\beta $-decay rates are extremely sensitive to both,  
 the asymptotics of the radial wave functions and to the correct accounting  
 for the dependence on the parent state.  
 First three columns with the results on $B$(GT) values in both tables are obtained  
 using formula (\ref{MGT5}), while  
 in the last column (denoted by a star) the radial dependence on  
 the parent state according to (\ref{MGT}) is taken into account.  
 The label IC(INC) refers to the isospin-conserving (isospin-nonconserving) interaction used,  
 and HO or WS indicate the type of the wave functions. Roman numbers I or II in Table II stand  
 for the method to fit the single-particle potentials as discussed in section  
 II. Only method II is used in the calculations for $sd$-shell nuclei.


\subsubsection{p-shell nuclei}  
  
 {\it A=8,9}.  
  
The analysis of light nuclei is often complicated from both theoretical and  
 experimental sides.  
 The studied $A=8$ triad includes the transition to the very broad $2^+$ state  
 of $^8$Be, unstable with respect to $\alpha \alpha$ breakup.  
 Moreover, in the description of the well-known proton-halo nucleus $^8$B, the  
 vicinity of the continuum can be important.  
 The decay of proton-rich $^9$C always continues through $2\alpha $-p three-body break-up,  
 which complicates the experimental analysis of corresponding transitions~\cite{A9}.  
 In addition, the specific feature of $^9$Li--$^9$C mirror pair  
 is the so-called anomalous isoscalar magnetic moment~\cite{muC9} which is still not fully  
 understood from a theoretical point of view.  
  

 The shell model yields good agreement of the energy spectra and  
 even nuclear moments with either harmonic oscillator or WS wave functions (see Table I)  
 for the lowest states of $A=8,9$, despite missing the width of the unbound states.  
 The only significant deviation is the quadrupole moment of the ground state of $^8$B,  
 showing necessity to go beyond the standard shell model we use.  
  
 However, as is clearly seen from Table II,  
 the shell model experiences difficulties in reproducing  
 the absolute values of the GT $\beta $-decay rates which are about ten and five times overestimated  
 for $A=8$ and $A=9$ systems, respectively.  
 The calculations with the best effective GT operator of Ref.~\cite{ChWa93} also overestimates  
 the transition rates in $A=8,9$, although with  
 a factor of two discrepancy.

 The apparent disagreement is unlikely to be the consequence  
 of a particular renormalization of the GT operator for these nuclei,  
 and is an indication of the limits of the theoretical approach used  
 (few-nucleon systems, clusterization effects, close continuum).  
 Following these arguments, one should not consider data  
on $A=8,9$ to extract second-class currents, 
until an accurate treatment by more fundamental microscopic techniques,  
 such as Green's Function Monte-Carlo methods~\cite{PiWi00},  
 and no-core shell model~\cite{NaBa98} is attained.  
 Note, that $A=8$ nuclei treated by the standard shell model  
 have mainly been used in the past  
 in the context of the second-class currents problem (e.g., ~\cite{Wi71,To73,Wi00}).  
  
  

 {\it A=12}.  
  
 The triad of $A=12$ nuclei, $^{12}{\rm B}(1^+;1)$ $\to $ $^{12}{\rm C}(0^+;0)$  
 and $^{12}{\rm N}(1^+;1)$ $\to$ $^{12}{\rm C}(0^+;0)$,  
 is the most famous and well studied case~\cite{A12}, since it was the first measured case of  
 large asymmetry and up to now with the best precision.  
 The electromagnetic moments for $1^+,T=1$ states in mother nuclei  
 are well reproduced.  
 The rates of the GT transitions to the ground state  
 of $^{12}$C are close to the measured ones, while the calculated $\beta $-decay rates  
 to the first excited state are about two times larger than the experimental values.  
 The use of effective GT-operator~\cite{ChWa93} helps to reduce the disagreement to only  
 20\% of overestimation.

 At present, there are already $ab$-initio shell-model calculations available  
 for $^{12}$C~\cite{C12}. It would be interesting to apply the model to calculate the GT  
 strength, since the questions such as limited configuration space and necessity of  
 auxiliary realistic potential are not encountered there.

 {\it A=13}.  
  
 The data on the rates for the GT transitions to the ground states $(\frac12^-, \frac12)$  
 in $A=13$ nuclei is also advantageous by its precision. The shell model well reproduces the  
 four lowest excited states of negative parity, including the values of the ground  
 state magnetic and quadrupole moments.  
 The positive parity states are outside of the $p$-shell model space.  
 The theoretical $\beta $-decay rates are a factor of 1.2 larger than the experimental ones.  
 No particular differences between wave functions obtained with the fitting procedure I and II  
 can be seen in this case.

 \subsubsection{sd-shell nuclei}  
  
 {\it A=17}.  
  
  We will devote a special discussion to $A=17$ nuclei, since they have already got  
  a lot of attention from  both experimental and theoretical sides, because of the enormously  
 large asymmetries of the first forbidden  
 $\beta $-decay transition rates~\cite{Borge,Mil97,Michel}.  
  
 Here we study these nuclei with respect to the asymmetries in the allowed GT decay modes,  
 which are also large.  
 Although the nuclei near the doubly magic ones are the most difficult  
 for the shell model framework, nuclei around $^{16}$O can be well described  
 within $p_{1/2}d_{5/2} s_{1/2}$-shell model space~\cite{ZBM}.  
 In the calculations we used the interaction  
 of Ref.~\cite{OrBr89}, which is an INC version of the interaction close  
 to the pioneer one~\cite{ZBM}.  
 Let us remark that the interaction exploited here gives much more realistic  
 spectra, in particular the positions of $3/2^-$ states in $^{17}$O and  
 $^{17}$F, and transition rates as compared to the $(psd)$-shell model interaction used in  
 Ref.~\cite{ChWa93}.  
  

 {\it A=20,25,28,31,35}.  
  
 As is well-known, the $sd$-shell nuclei are fairly good described  
 by the interaction of Wildenthal~\cite{usd,BrWi88},  
 while the INC corrections take care about the subtle effects.  
 The systematic calculations of $\beta $-decay rates in $sd$-shell nuclei have been already  
 performed~\cite{BrWi85,BrWi88}, although without accounting for INC effects, and are known to be in  
 a good agreement with the experiment.  
  
 The results on $B$(GT)-values for $sd$-shell nuclei, obtained with both IC and INC Hamiltonians and  
 with harmonic oscillator or WS radial wave functions are summarized in Table III.  
 There are still no data on the transitions for $A=21$ and $A=25$ nuclei, however, the measurements are  
 under way~\cite{proposal}.  
 The calculated $B$(GT) values are all about a factor of $0.5 - 0.7$ quenched compared  
 to the experiment, except for the $A=20$ case ($T=2 \to T=1$ transitions).  
  
 As in the case of $p$-shell nuclei, there was no principal  
 difference between the radial wave functions, however, the values obtained with  
 formula (\ref{MGT}) are assumed to be the most realistic ones.  
 We did not take into  
 account the dependence on the parent states for $A=25,28,31$  
 because of the extremely large dimensions of the model spaces, as well as for  
 $A=35$ (S-CL-K-Ar) for which the experimental error bars are very large.

 \subsection{Asymmetries of GT transition rates}  
  
 We have performed, for the first time, a detailed study of the asymmetries for a selected  
 ensemble of $p$- and $sd$-shell nuclei. The chosen nuclei are on one hand rather  
 well described by the shell model and on the other hand the corresponding  
 measured asymmetries present small error bars (Table IV and Fig.~4).  
  

We have calculated the asymmetries of the $B$(GT)-values in the mirror nuclei  
 due either to the use of the isospin non-conserving (INC) interaction or/and  
 to differences in the tails of wave functions of the decaying proton or neutron.  
 We present the effect of the isospin non-conserving interaction alone (INC+HO),  
 of the mismatch of the radial integrals alone  
 (IC+WS) and of both (INC+WS). In the latter case, we show  
 the influence of the inclusion of the dependence on the parent states  
 in the radial wave functions ((INC+WS)$^*$).  
  
We also present the first estimates of the asymmetries for some $sd$-shell nuclei  
for which the shell model description is rather good, but the asymmetries either present too  
large error bars or their analysis is under study \cite{comp} (Table V and Fig.~4).  
  
 All the results shown in Tables IV and V correspond to the radial wave functions obtained with method II  
 by varying the volume term of the potential in the fitting procedure.  
 We have checked that fitting the single-particle energies with the  
 inclusion of a surface term changes the results by a few percent.  
 We have also calculated the asymmetries using method I.  
 This always leads to much larger asymmetries than those determined by  
 method II and, in particular, it gives unrealistic ones for cases where one  
 single-particle state is unbound, such as for $A=8,9$.  
 We have also mentioned above that it leads to unrealistic results for the quadrupole moments of  
 some light nuclei. For these  
 reasons and because of a better consistency with the shell model,  
 we believe that method II (and not method I) should be used in the procedure of fitting the wave functions  
 contrary to the previous studies~\cite{Wi71,To73}.

 As we can see from Table IV,  
 the effect due to the differences in the tails of the wave functions  
 is by far the largest in the $p$-shell nuclei; whereas in the case of  
 the $sd$-shell nuclei, it can be sometimes of the same order as the  
 asymmetry due to the INC interaction. This happens when there is  
 a large contribution of the $d_{5/2}$ or $d_{3/2}$ single-particle states for which  
 the centrifugal barrier tends to confine the wave function inside  
 the potential reducing the differences in the tails.  
 The case $A=17$, including the halo nucleus $^{17}$F, represents an  
 exception : the asymmetry due to the INC interaction is  
 almost ten times larger than the one due to the mismatch of the  
 wave functions.  
  
 Concerning the dependence on the parent states (Eq.~(\ref{MGT})),  
 we see from Table IV that its inclusion has a significant effect on the asymmetries. As it was found in  
 \cite{Wi71,To73}, the use of Eq.(\ref{MGT}) instead of Eq.(\ref{rad}) may significantly increase or  
 decrease the asymmetry and in a few cases up to a factor of two.  
  
We have compared the theoretical asymmetries to the ones obtained in Ref.~\cite{Wi71,To73},  
 although those calculations differ from ours in the parametrization of the shell model Hamiltonian,  
 the WS potential, as well as in the fitting procedure (method I was used).  
 In general, our asymmetries  are much  larger than previously calculated in Ref.~\cite{Wi71,To73}.  
 However, for $A=12$ case, our result is in reasonable agreement with their values,  
 as well as with the value in \cite{Bl71} where method II was used and also a very similar parametrization  
 of the WS potential.

As far as the comparison with the measured asymmetries is concerned (Table IV),  
 we see that our calculated asymmetries due to the only approximate isospin symmetry in the nucleus  
 (described in our case through the use of the INC interaction \cite{OrBr89} and of WS  
 radial wave functions) : a) are compatible with the measured asymmetries in  
 a few cases; b) overestimate the measured values for  
 $A=8,12(g.s.),17$;  
 c)  underestimate the experiments for $A=9,20$ (Fig.~4).  
 Remark that $A=8,17$ represent rather delicate cases because they  
 include halo nuclei which should be treated with particular techniques.  
 For $A=8,9$, the theoretical $B$(GT)-values are larger than the experimental  
 ones by a factor of four.  
  
  For $p$-shell nuclei, these conclusions do not change even  
 after the inclusion of the other above mentioned higher order corrections to $\delta^{nucl}$,  
 which have been found in Ref.~\cite{To73} not to exceed 1\%.

\section{Conclusions}  
  
 Within the shell model,  
 we have calculated the Gamow-Teller $\beta_+ $ and $\beta_-$ decay rates for  
 an ensemble of light $p$- and $sd$-shell mirror nuclei. The theoretical results are  
 important in connection with the possible existence  
 of second-class currents in the weak interaction.  
 To assess their existence, it is necessary to carry out accurate estimates  
 of the asymmetries coming from nuclear structure aspects.  
  
The calculations are performed systematically  for the known experimental cases among $p$-shell and,  
 for the first time, $sd$-shell nuclei.  
 We have evaluated the asymmetry due to the only  
 approximate isospin symmetry of the shell model Hamiltonian  
 and differences between the proton and neutron radial wave functions.  
 The former is accounted for through the use of an isospin non-conserving  
 interaction which describes well the shifts in energy levels of the mirror nuclei.  
 The latter is due to the use of realistic single-particle radial wave functions from  
 a WS potential adjusted to reproduce the experimental proton and neutron separation energies.  
  
 We have thoroughly studied the sensitivity of the asymmetries  
 to the inclusion of the parent state dependence as well as  
 to different fitting procedures, namely by fitting either the energy of the last  
 unoccupied state or the energies of all the single-particle states in the shell  
 at the same experimental separation energy. We argue that, contrary to what was  
 done in the past, the latter fitting procedure is more consistent than the former.

The theoretical asymmetries we obtain are on average larger than all previous estimates.  
 The results are compatible with the measured values in a few cases.  
 For $A=8,12,17$, the theoretical values are larger  
 than the experimental ones, whereas for $A=9,20$ we find the opposite.  
  
As is clear from the present study,  
 the average reliability of $B$(GT)-values,  
 which determines the level of confidence for the theoretical asymmetries,  
 seems still not to be high enough for  
 extracting the second-class currents from  the differences  
 in nuclear GT $\beta $-decay rates, even  
 at the present level of many-body techniques.  
 
 In this work, we restricted ourselves to WS wave functions.  
 Calculations with Hartree-Fock wave functions  
 would not modify the quality of the $B$(GT)-values significantly enough to change these conclusions. 
 Nevertheless, these calculations would be needed, once a 
 better overall description of the transition rates is attained. 
 
 More extensive and systematic data would be very helpful.

\vspace{1cm}  
  
\acknowledgements{ We thank Bertram Blank and Jean-Charles Thomas for  
 making us aware of the problem,  
 providing us with the systematics of available experimental data prior to publication  
 and stimulating interest to our work at earlier stages.  
 We are grateful to Nicole Vinh Mau for her kind help at the beginning of our work and the very  
  interesting numerous discussions.  
  We gratefully acknowledge the discussions, also in e-mail correspondence,  
  with H.~Sazdjian and E.~Ormand.  
 N.A.S. thanks IPN Orsay for hospitality and CNRS, France,  
 for visitor support, and acknowledges the Institute for Nuclear Theory at the University of Washington  
 for hospitality and  the U.S. Department of Energy for partial support during  
 the completion of this work. N.A.S. thanks K.~Heyde for valuable comments.
 Financial support of the DWTC (grant IUAP \#P5$/$07) is acknowledged.}
  

\begin{table}  
\label{moments}  
 \caption{Comparison of experimental and theoretical static magnetic dipole ($\mu$) in $\mu_N$ and  
 electric quadrupole ($Q$) in e.fm$^2$ moments of the ground states of the $p$-shell  
 nuclei of interest.  
 The quadrupole moments have been calculated either with harmonic oscillator (HO) or Woods-Saxon  
 (WS) wave functions using method II with a volume term (see text), according to (\protect\ref{em}).  
 The experimental data are taken from Ref.~\protect\cite{Raghavan}, if not specified.}  
\begin{tabular}{llccccccccc}  
 \hline  
 & $J^{\pi };T $ & $\mu $ & $\mu $  
 & $Q$(HO)  & $Q$(WS)  & $Q$(HO)  & $Q$(WS)  & $Q$(HO)  & $Q$(WS)   & $Q$  \\  
 & &  \multicolumn{1}{c}{ g$_s^{free}$ }  
 & \multicolumn{1}{c}{ EXP }  
 &  \multicolumn{2}{c}{ ($e_{\pi }=1.5, \, e_{\nu }=0.5$)}  
 &  \multicolumn{2}{c}{ ($e_{\pi }=1.35, \, e_{\nu }=0.5$)}  
 &  \multicolumn{2}{c}{ ($e_{\pi }=1.35, \, e_{\nu }=0.35$)}  
 &  \multicolumn{1}{c}{EXP }\\  
\hline  
 $^8$Li & $2^+$;1  & 1.372 & 1.654 & 2.35 & 3.99 & 2.22 & 3.86 & 1.87 & 3.05 & 3.17(4)\\  
 $^8$B  & $2^+$;1   & 1.245 & 1.036 & 3.83 & 5.24 & 3.48 & 4.84 & 3.36 & 4.46 & 6.83(21)\protect\cite{QB8}\\  
\hline  
 $^9$Li & $\frac32^-$;$\frac32$ & 3.488 &  
 3.4335(52) & $-4.23$ & $-5.14$ & $-3.89$ & $-4.78$ & $-3.62$ & $-4.32$ & 2.78(8)\\  
 $^9$Be & $\frac32^-$;$\frac32$ & $-1.229$  
 & $-1.1778(9)$  & 4.59 & 5.80 & 4.20 & 5.43 & 3.96 & 4.81 & $+5.3(3)$ \\  
 $^9$C & $\frac32^-$;$\frac32$ & $-1.5595$ &  
 ~~$(-)1.3914(5)$\protect\cite{muC9}~~  & & & & & & & \\  
\hline  
 $^{12}$B & $1^+$;1   & 0.767 & $+1.003$ & 2.18 & 2.42 & 1.97 & 2.20 & 1.93 & 2.14 & 1.34(14)\\  
 $^{12}$N  & $1^+$;1   & 0.570 & $+0.457$ & 1.013 & 1.50 & 0.976 & 1.42 & 0.77 & 1.21  
 & $+0.98(9)$\protect\cite{QN12}\\  
\hline  
 $^{13}$B & $\frac32^-$;$\frac32$ &  
   3.147 & $+3.1778(5)$ & 4.73 & 4.89 & 4.24 & 4.40 & 4.24 & 4.40 & 3.74(40) \\  
 $^{13}$C & $\frac12^-$;$\frac12$ & 0.748 & $+0.7024$ & & &  &  &  &  &  \\  
 $^{13}$N & $\frac12^-$;$\frac12$ &  $-0.380$ & $(-)0.322$ &  &  &  &  &  &  &  \\  
 $^{13}$O & $\frac32^-$;$\frac32$ & $-1.378$ & $-1.3891(3)$\protect\cite{muO13}  
 & 1.56 & 1.61 & 1.56 & 1.61 & 1.09 & 1.13 & 1.10(13)\protect\cite{QO13}\\  
\hline  
\end{tabular}  
\end{table}

\begin{table}  
 \label{pshell}  
 \caption{Comparison of experimental and theoretical $B$(GT)-values in  
 $p$-shell mirror nuclei.  
 The results shown are obtained with the isospin-nonconserving (INC)  
 interaction and either harmonic oscillator (HO) or Woods-Saxon (WS) wave functions   
 determined with method I or II (see text) by adjusting the volume term,  
 without including the dependence on the parent state.  
 The results obtained by including a surface term are shown into parenthesis.  
 The $B$(GT)-values given in column (INC+WS)$^*$ incorporate the dependence on the parent states as  
 follows from Eq.~(\protect\ref{MGT}). The experimental values~\protect\cite{comp} are obtained  
 following the procedure of Refs.~\protect\cite{BrWi85,ChWa93}   
 on the basis of $Q$-values from Ref.~\protect\cite{Audi}, halflives and  
 branching ratios  from Ref.~\protect\cite{ChWa93}, if not specified.}  
\begin{tabular}{lllcccccc}  
 \hline  
 &  $J^{\pi }_i;T_i$ & $J^{\pi }_f;T_f$  & INC+HO &  I:~INC+WS &  II:~INC+WS & II:~(INC+WS)* & EXP \\  
\hline  
 $^8$Li$(\beta^-)$$^8$Be & $2^+$;1 & $2^+_1$;0  & 0.1170  & 0.0265(0.0295)  
 & 0.1006 (0.1001) & 0.0984 & 0.0158(2) \\  
 $^8$B$(\beta^+)$$^8$Be & $2^+$;1 & $2^+_1$;0 & 0.1135  & 0.0082(0.0092)  
 &  0.0863 (0.0864) & 0.0828 & 0.0146(2) \\  
\hline  
   $^9$Li$(\beta^-)$$^9$Be & $\frac32^-;\frac32$ &  $\frac32^-_1;\frac32$ (g.s.) &  0.1498 &  
    0.1052 (0.1101) & 0.1375 (0.1352) &  0.1417 & 0.0296(18)~\cite{Nym} \\  
 $^9$C$(\beta^+)$$^9$B & $\frac32^-;\frac32$ & $\frac32^-_1;\frac32$ (g.s.) & 0.1517 & 0.0740 (0.0778)  
 &  0.1310 (0.1283) &  0.1379 & 0.0256(3)~\cite{A9} \\  
\hline  
 $^{12}$B$(\beta^-)$$^{12}$C & $1^+$;1 & $0^+_1$;0 (g.s.) & 0.5078  & 0.4250 (0.4528)  
 & 0.4703 (0.4692) &  0.4250 & 0.5272(18) \\  
 &  & $2^+_1$;0   & 0.1038  & 0.0979 (0.0999)  
 & 0.0990 (0.0984) &  0.0954 & 0.0476(2) \\  
 $^{12}$N$(\beta^+)$$^{12}$C & $1^+$;1 & $0^+_1$;0 (g.s.)  & 0.5045  & 0.3629 (0.3918)  
 & 0.4272 (0.4245) &  0.3488 & 0.4682(31) \\  
 &  & $2^+_1$;0   & 0.1045  & 0.0875 (0.0897)  
 & 0.0917 (0.0909) & 0.0857 & 0.0435(8)   \\  
\hline  
 $^{13}$B$(\beta^-)$$^{13}$C & $\frac32^-;\frac32$ & $\frac12^-_1;\frac12$ (g.s.) & 0.767 &  
    0.6861 (0.7096) & 0.7206 (0.7218) &  0.7181 & 0.5671(72)\\  
 $^{13}$O$(\beta^+)$$^{13}$N & $\frac32^-;\frac32$ & $\frac12^-_1;\frac12$ (g.s.) &  0.752 &  
    0.6095 (0.6381) & 0.6580 (0.6615) &  0.6609 & 0.5097(130) \\  
 \hline  
\end{tabular}  
\end{table}

\begin{table}  
 \label{sdshell}  
 \caption{Comparison of experimental and theoretical $B$(GT)-values in  
   $sd$-shell mirror nuclei.  
 We present results obtained with the isospin conserving (IC) and isospin non-conserving (INC)  
 interactions and either  harmonic oscillator (HO) or Woods-Saxon (WS) wave functions   
 determined from method II by adjusting a volume term.  
 The results including the dependence on the parent state are shown in column (INC+WS)$^*$  
 only for the nuclei considered in the discussion of the asymmetries.  
 The experimental values~\protect\cite{comp} are obtained  
 following the procedure of Refs.~\protect\cite{BrWi85,ChWa93}   
 on the basis of $Q$-values from Ref.~\protect\cite{Audi}, halflives and  
 branching ratios as specified.}  
\begin{tabular}{lllccccc}  
 \hline  
 & $J^{\pi }_i;T_i$ & $J^{\pi }_f;T_f$  & INC+HO  & IC+WS  & INC+WS & (INC+WS)$^*$ & EXP \\  
\hline  
 $^{17}$N$(\beta^-)$$^{17}$O & $\frac12^-;\frac32$ & $\frac32^-_1;\frac12$  & 0.2079 &  
  0.2333 & 0.2021 & 0.2023 & 0.2342(89)~\cite{Tilley} \\  
 $^{17}$Ne$(\beta^+)$$^{17}$F & $\frac12^-;\frac32$ & $\frac32^-_1;\frac12$  & 0.1356 &  
  0.2216 & 0.1272 & 0.1228 & 0.1632(50)~\cite{Tilley,Dom94}\\  
\hline  
 $^{20}$F$(\beta^-)$$^{20}$Ne & $2^+;1$ & $2^+_1;0$ & 0.0921 & 0.0948 & 0.0922  
 & 0.0924 & 0.0645(1)~\cite{Tilley} \\  
 $^{20}$Na$(\beta^+)$$^{20}$Ne & $2^+;1$ & $2^+_1;0$  & 0.0935 & 0.1098 &  
 0.1085 & 0.0914 & 0.0633(4)~\cite{Tilley}\\  
\hline  
 $^{20}$O$(\beta^-)$$^{20}$F & $0^+;2$ & $1^+_1;1$  & 0.9124 & 0.9571 & 0.9183  
 & 0.9031 & 1.1166(48)~\cite{Tilley} \\  
 $^{20}$Mg$(\beta^+)$$^{20}$Na & $0^+;2$ & $1^+_1;1$ & 0.9120 & 0.9439 & 0.9054 & 0.9171 &  
 0.9445(666)~\cite{Audi,Tilley}  \\  
  \hline  
 $^{21}$F$(\beta^-)$$^{21}$Ne & $\frac52^+;\frac32$ & $\frac32^+_1;\frac12$ (g.s.) & 0.0215 & 0.0226 & 0.0215 & & \\  
 &  & $\frac52^+_1;\frac12$ & 0.1188 & 0.1243 & 0.1195 & & \\  
 &  & $\frac72^+_1;\frac12$ & 0.1159 & 0.1165 & 0.1168 & & \\  
 $^{21}$Mg$(\beta^+)$$^{21}$Na & $\frac52^+;\frac32$ & $\frac32^+_1;\frac12$  (g.s.) & 0.0224 & 0.0218 & 0.0216 & & \\  
 &  & $\frac52^+_1;\frac12$ & 0.1209 & 0.1214 & 0.1188 & & \\  
 &  & $\frac72^+_1;\frac12$ & 0.1128 & 0.1149 & 0.1120 & & \\  
 \hline  
 $^{25}$Na$(\beta^-)$$^{25}$Mg & $\frac52^+;\frac32$ & $\frac52^+_1;\frac12$  (g.s.)  &  
 0.0458 & 0.0472 & 0.0463 & & \\  
 & & $\frac32^+_1;\frac12$ & 0.0836 & 0.0820 & 0.0830 & & \\  
 &  & $\frac72^+_1;\frac12$ & 0.0634 & 0.0615 & 0.0643 & & \\  
 & & $\frac52^+_2;\frac12$ & 0.0048 & 0.0052 & 0.0047 & & \\  
 $^{25}$Si$(\beta^+)$$^{25}$Al & $\frac52^+;\frac32$ & $\frac52^+_1;\frac12$  (g.s.)  &  
 0.0463 & 0.0462 & 0.0458 & & \\  
 &  & $\frac32^+_1;\frac12$ & 0.0765 & 0.0798 & 0.0738 & & \\  
 &  & $\frac72^+_1;\frac12$ & 0.0589 & 0.0595 & 0.0578 & & \\  
 &  & $\frac52^+_2;\frac12$ & 0.0049 & 0.0053 & 0.0050 & & \\  
\hline  
 $^{28}$Al$(\beta^-)$$^{28}$Si & $3^+;1$ & $2^+_1;0$  & 0.1458 & 0.1444 & 0.1435 & & 0.0825(1)~\cite{Endt}\\  
 $^{28}$P$(\beta^+)$$^{28}$Si & $3^+;1$ & $2^+_1;0$  & 0.1449 & 0.1303 & 0.1287 & & 0.0870(9)~\cite{Endt}\\  
\hline  
 $^{31}$Al$(\beta^-)$$^{31}$Si & $\frac52^+;\frac52$ & $\frac32^+_1;\frac32$  (g.s.)  & 0.2120 &  
  0.2090 & 0.2124 & & 0.1049(122)~\cite{Endt} \\  
 $^{31}$Ar$(\beta^+)$$^{31}$Cl & $\frac52^+;\frac52$ & $\frac32^+_1;\frac32$  (g.s.)  & 0.1947 &  
  0.1958 & 0.1827 & & 0.0788(53)~\cite{Fynbo} \\  
\hline  
 $^{35}$S$(\beta^-)$$^{35}$Cl & $\frac32^+;\frac32$ & $\frac32^+_1;\frac12$  (g.s.)  & 0.0861 & 0.0870  
 & 0.0865 & & 0.0588(2)~\cite{Endt}\\  
 $^{35}$K$(\beta^+)$$^{35}$Ar & $\frac32^+;\frac32$ & $\frac32^+_1;\frac12$  (g.s.)  & 0.0846 & 0.0843  
 & 0.0824 & & 0.0496(84)~\cite{Endt}\\  
 \hline  
 $^{35}$P$(\beta^-)$$^{35}$S & $\frac12^+;\frac52$ & $\frac12^+_1;\frac32$  & 0.9052 & 0.8572  
 & 0.8836 & 0.8861 & 0.4591(70)~\cite{Trin}\\  
 $^{35}$Ca$(\beta^+)$$^{35}$K & $\frac12^+;\frac52$ & $\frac12^+_1;\frac32$  & 0.8572 & 0.7435  
 & 0.7257 & 0.7458 & 0.3891(154)~\cite{Trin}\\  
 \hline  
\end{tabular}  
\end{table}  
  
\begin{table}  
 \label{asymmetryfin1}  
 \caption{Comparison of experimental and theoretical  
 asymmetries  $\delta $ ($\%$) of B(GT)-values for transitions in  
 selected $p-$ and $sd$-shell mirror nuclei.  
 Entries denoted $A=20$, $A=20$(a) and $A=35$ correspond to  O-F-Mg-Na, F-Ne-Na and  
 P-S-Ca-K systems, respectively (see Table II for the details).  
 The different columns give the asymmetries obtained with the  
 isospin conserving (IC) or isospin non-conserving (INC)  
 interactions and harmonic oscillator (HO) or Woods-Saxon (WS) wave  
 functions obtained with method II (see text), adjusting a volume term.  
 The column (INC+WS)$^{*}$ shows  the results obtained with the inclusion of the dependence on the parent  
 state.  The experimental asymmetries are calculated on the basis of the data  
 from Tables II - III (see also are Ref.~\protect\cite{comp}).}  
\begin{tabular}{lllcccccc}  
 \hline  
 $A$ & $J^{\pi }_i;T_i$ & $J^{\pi }_f;T_f $ &  
 INC+HO & IC+WS &INC+WS &  (INC+WS)$^{*}$ & EXP. \\  
 \hline $8$ & $2^+$;1 &  $2^+_1$;0 &  3.10 &  13.21 & 16.65 & 18.82   &  
 $8.4 \pm 1.8$  \\  
  $9$ & $\frac32^-;\frac32$ & $\frac32^-_1;\frac12$ (g.s.) & $-1.26$ & 6.4  & 4.95 &  
  2.72  & $16 \pm 8$ \\  
 $12$ &  $1^+$;1 & $0^+_1$;0 (g.s.) & 0.70  & 9.4 & 10.09  & 21.85  & $12.6 \pm 0.8$ \\  
 $$ & $$ & $2^+_1$;0 & $-0.65$ & 8.45 & 7.92 & 11.33  & $9.5\pm 1.9$ \\  
 $13$ &  $\frac32^-;\frac32$ &   $\frac12^-_1;\frac12$ (g.s.) & 2.10 & 7.26 & 9.50  
  & 8.65  & $11.3 \pm 3.2$ \\ \hline  
 $17$&  $\frac12^-;\frac32$ & $\frac32^-_1;\frac12$  & 53.26 & 5.29 &  
 58.90 & 64.80  & $44\pm 7$ \\  
 $20$  & $0^+$;2 & $1^+_1$;1 & 0.04 & 1.41 & 1.42  & $-1.53$   & $18\pm 8$ \\  
 $20$(a)  & $2^+$;1 & $2^+_1$;0 &  $-1.45$ & $-13.64$ & $-14.99$ & 1.11 & $1.8  
 \pm 0.7$ \\  
 $35$ & $\frac12^+;\frac52$ & $\frac12^+_1;\frac32$ &  5.59 & 15.28 &  
 21.75 & 18.81 & $18 \pm 5$ \\  
 \hline  
\end{tabular}  
\end{table}  
  
\begin{table}  
 \label{asymmetryfin2}  
\caption{Comparison of experimental and theoretical  
 asymmetries  $\delta $ ($\%$) of B(GT)-values for transitions in  
  $sd$-shell mirror nuclei.  
 The entry denoted as $A=35$ correspond to the S-Cl-K-Ar system (see Table II for the details).  
 The different columns give the asymmetries obtained with  
 the isospin conserving (IC) and isospin non-conserving (INC)  
 interactions with the harmonic oscillator (HO) or Woods-Saxon (WS) wave  
 functions obtained with method II (see text), adjusting a volume term.   
 The experimental asymmetries are calculated on the basis of the data  
 from Tables II - III (see also are Ref.~\protect\cite{comp}). }  
\begin{tabular}{lllcccc}  
 \hline  
 $A$ & $J^{\pi }_i;T_i$ & $J^{\pi }_f;T_f $ &  INC+HO & IC+WS  
 & INC+WS & EXP. \\  
\hline  
 $21$&  $\frac52^+;\frac32$ & $\frac32^+_1;\frac12$ (g.s.) &  $-3.96$ & 3.49  
 &  $-0.56$ &   \\  
  &$$ &  $\frac52^+_1;\frac12$ & $-1.71$  & 2.35 & 0.60  &   \\  
 & $$ & $\frac72^+_1;\frac12$  & 2.78 & 1.44 & 4.23  &  \\  
 $25$ &  $\frac52^+;\frac32$ & $\frac52^+_1;\frac12$ (g.s.) & $-1.09$ & 2.24 & 1.11 &   \\  
  & $$ &  $\frac32^+_1;\frac12$ & 9.21 & 2.83 & 12.39 &   \\  
 &$$ & $\frac72^+_1;\frac12$  & 7.76 & 3.83 & 11.23 &  \\  
 &$$ & $\frac52^+_2;\frac12$  & $-3.17$ & $-2.52$  & $-5.58$ &  \\  
 $28$ &  $3^+$;1 & $2^+_1$;0 & 0.61 & 10.85 & 11.54 &  $-5 \pm 1$  \\  
 $31$ & $\frac52^+;\frac52$ &  $\frac32^+_1;\frac32$ (g.s.) & 8.88 & 6.72  
 & 16.27 &  $33\pm 18$ \\  
 $35$  &  $\frac32^+;\frac32$ & $\frac32^+_1;\frac12$ (g.s.) & 1.82 & 3.13  
 & 4.96 & $18.4\pm 20.0$ \\  
 \hline  
 \end{tabular}  
 \end{table}

 \begin{figure}[htbp]  
 \begin{center}  
  \includegraphics[width=10cm]{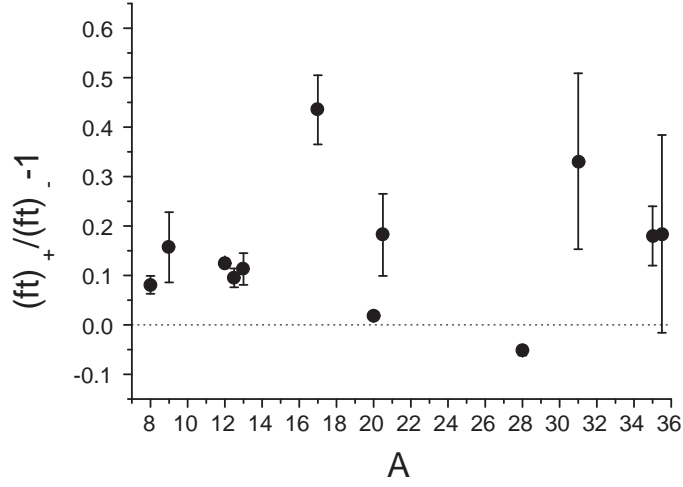}  
 \caption{Experimental asymmetry (\protect\ref{delta1}) deduced from the data on  
 $\beta_{\pm }$ mirror transitions characterized by $|\Delta T|=1$, $|\Delta J|=0,1$,  
 $\pi _i=\pi _f$ (calculated on the basis of the experimental $B$(GT)-values given in Tables II  
 and III; see the captions of the tables for references).   
 The results for the second transitions in  
 $A=12$, $A=20$ O-F-Mg-Na and $A=35$ S-Cl-K-Ar are shifted by 0.5 unit  
 from the $A$-value.}  
 \end{center}  
 \end{figure}  
  
\begin{figure}[htbp]  
 \begin{center}  
 \includegraphics[width=70mm]{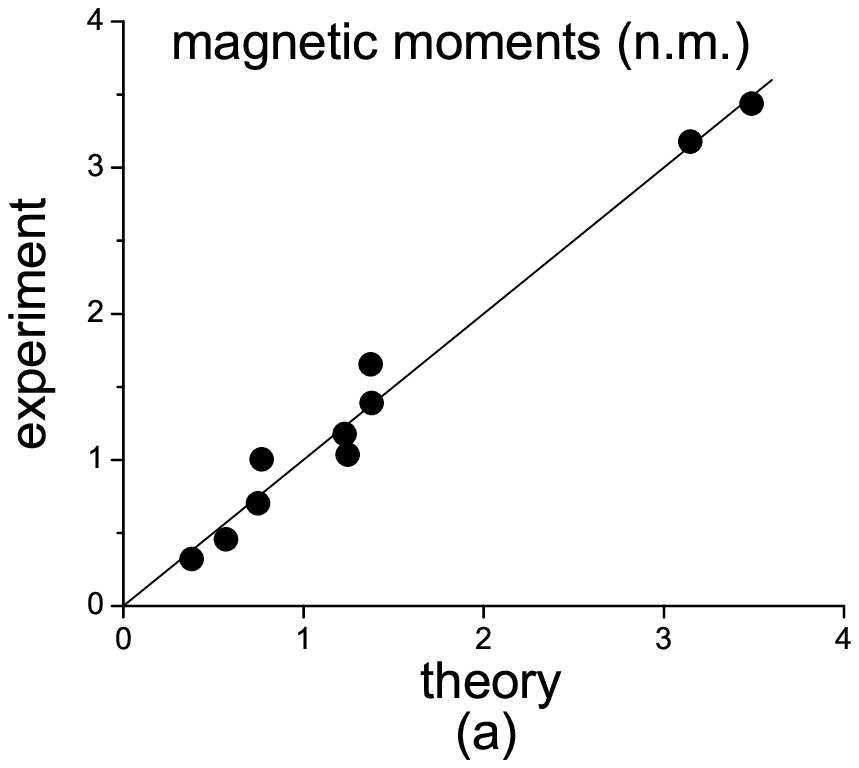}  
 \includegraphics[width=70mm]{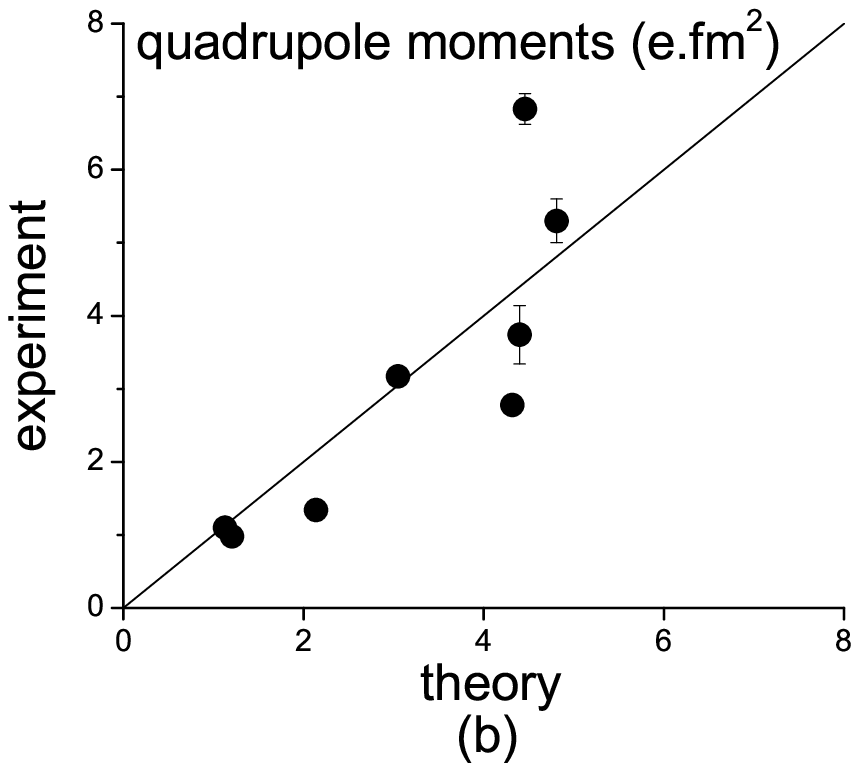}  
 \caption{Experimental versus theoretical (Table I): (a) absolute values of magnetic dipole moments  
 of the ground states of $p$-shell nuclei obtained with the free-nucleon $g$-factors  
 (data  are taken from Ref.~\protect\cite{Raghavan,muC9,muO13});  
 (b) absolute values of electric quadrupole moments of the ground states  
 of $p$-shell nuclei obtained with the WS wave functions and  
 $e(\pi )=1.35$, $e(\nu )=0.35$ (data  are taken from Ref.~\protect\cite{Raghavan,QB8,QN12,QO13}).  
 Solid lines correspond to the ratios equal to 1 and are given to guide the  
 eye.}  
 \end{center}  
 \end{figure}

 \begin{figure}[htbp]  
 \begin{center}  
\includegraphics[width=80mm]{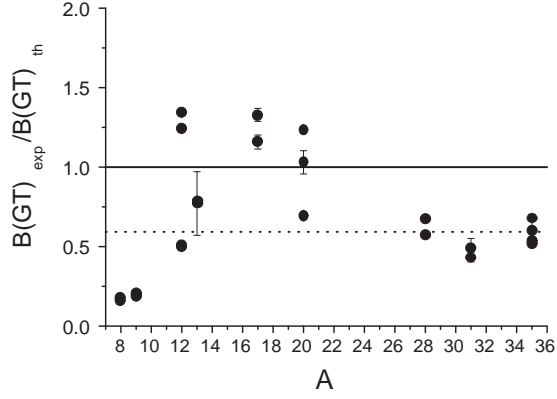}  
 \caption{Ratios of experimental to theoretical $B$(GT) values present in Tables II and III  
 (see the captions of the tables for  references on the experimental data).   
 Theoretical values are those  
 obtained  with the dependence on the parent state (column (INC+WS)* in Tables  
 II and III) and without dependence for  
 $A=28,31$ and $A=35$ S-CL-K-Ar (column (INC+WS) in Table III)          .  
 Solid and dotted lines correspond to the rations 1 and $(0.77)^2$, respectively, and are given  
 to guide the eye.}  
 \end{center}  
 \end{figure}  
  
\begin{figure}  
\begin{center}  
 \includegraphics[width=80mm]{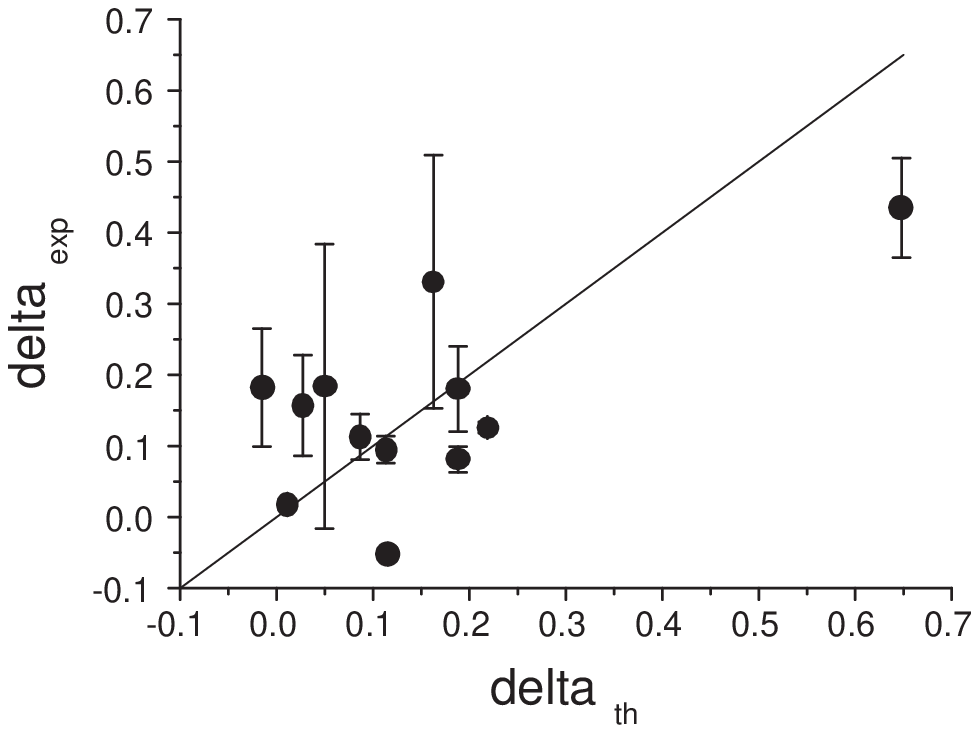}  
 \includegraphics[width=80mm]{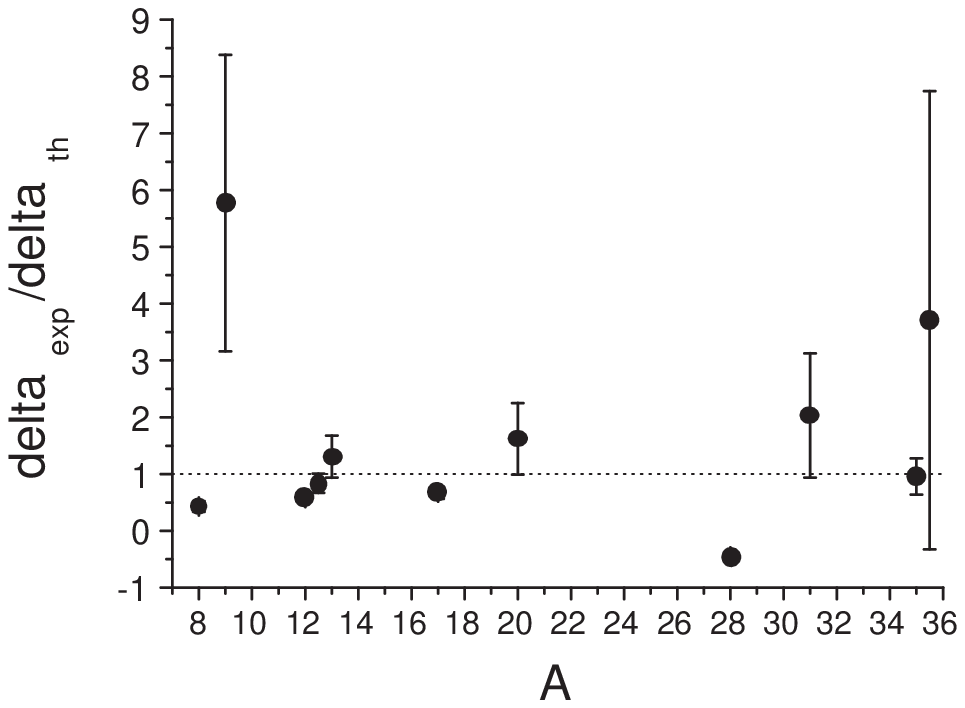}  
\end{center}  
\protect\caption{Experimental versus theoretical asymmetries $\delta $ (left)  
  and ratios of experimental to theoretical  
  asymmetries (right) of selected GT decay rates for $p$- and $sd$-shell nuclei. For a  
  given mass, the theoretical points show the effect due to the inclusion of  
  the dependence on the parent state, except for $A=28,31$ and $A=35$ S-Cl-K-Ar system.  
 The results for the second transitions in  
 $A=12$ and $A=35$ S-Cl-K-Ar are shifted by 0.5 unit from the $A$-value. The large in absolute value  
 ratio of the asymmetries for $A=20$ case O-F-Mg-Na is excluded.  
 The experimental asymmetries are calculated on the basis of the experimental  
  $B$(GT)-values from Tables II and III (see the captions of the tables for references).}  
\end{figure}

\end{document}